\documentstyle[12pt,blois,epsfig]{article}
\def\HI{\hbox{H~$\scriptstyle\rm I\ $}}

\def\nH{{\rm H}}

\def\kms{\,{\rm km\,s^{-1}}}
\def\kmsmpc{\,{\rm km\,s^{-1}\,Mpc^{-1}}}

\def\eblunits{\,{\rm nW\,m^{-2}\,sr^{-1}}}

\def\ndotunits{\,{\rm s^{-1}\,Mpc^{-3}}}

\def\msun{\,{\rm M_\odot}}
\def\mden{\,{\rm M_\odot\,Mpc^{-3}}}
\def\sfrd{\,{\rm M_\odot\,yr^{-1}\,Mpc^{-3}}}
\def\sfr{\,{\rm M_\odot\,yr^{-1}}}

\def\Lya{Ly$\alpha\ $}
\def\etal{{et al.\ }}
\def\AB{{\rm AB}}
\def\spose#1{\hbox to 0pt{#1\hss}}
\def\lta{\mathrel{\spose{\lower 3pt\hbox{$\mathchar"218$}}
     \raise 2.0pt\hbox{$\mathchar"13C$}}}
\def\gta{\mathrel{\spose{\lower 3pt\hbox{$\mathchar"218$}}
     \raise 2.0pt\hbox{$\mathchar"13E$}}}

\begin{document}
\heading{After the dark ages: the evolution of \\ luminous sources at ${\bf z<5}$}
\vspace{1cm}
\begin{center}
Piero Madau\\
{\it Space Telescope Science Institute, Baltimore, USA.}
\end{center}
\vspace{1cm}
\begin{bloisabstract}

I review recent observational and theoretical progress in our understanding 
of the cosmic evolution of luminous sources. Through a combination of deep 
{\it HST} imaging,  
Keck spectroscopy, and {\it COBE} background measurements, important 
constraints have emerged on the emission history of the galaxy population as 
a whole. A simple stellar evolution model,  
defined by a star-formation density that rises from $z=0$ to $z\approx 1.5$,
a universal Salpeter IMF, and a moderate amount of dust with $A_V=0.23$ 
mag ($A_{1500}=1.2$ mag), is able to account for most of the optical-FIR 
extragalactic background light, and reproduces the global ultraviolet, 
optical, and near-IR photometric properties of the universe. By contrast, 
a star-formation
density that stayed roughly constant at all epochs appears to overproduce 
the local $K$-band luminosity density. While the bulk of the stars 
present today formed relatively recently, the 
existence of a decline in the star-formation density above $z\approx 2$ 
remains uncertain. If stellar sources are responsible for photoionizing the 
intergalactic medium at $z\approx 5$, the rate of star formation at this 
epoch must be comparable or greater than the one inferred from optical 
observations of galaxies at $z\approx 3$. A population of dusty AGNs 
at $z\lta 2$ could make a significant contribution 
to the FIR background if the accretion efficiency is $\sim 10\%$.

\end{bloisabstract}

\section{Introduction}

In the last few years, the remarkable progress in our understanding of faint 
galaxy data made possible by the combination of HST deep imaging \cite{W96} 
and ground-based spectroscopy \cite{Li96}, \cite{El96}, \cite {S96} has 
permitted to shed some light on the evolution of the stellar birthrate in the
universe, to tentatively identify the epoch $1\lta z\lta 2$ where most of 
the optical extragalactic background light was produced, and to set important
contraints on galaxy evolution scenarios \cite{M98}, \cite{S98}, \cite{Bau98},
\cite{Gui97}. The 
explosion in the quantity of information available on the high-redshift 
universe at optical wavelengths has been complemented by the detection of 
the far-IR/sub-mm background by DIRBE and FIRAS \cite{Ha98}, \cite{Fix98}, 
that has revelead the optically `hidden' side of galaxy formation, and 
shown that a significant fraction of the energy released by stellar 
nucleosynthesis is re-emitted as thermal radiation by dust. The underlying 
goal of all these efforts is to understand the growth of cosmic structures
and the mechanisms that shaped the Hubble sequence, and ultimately to map   
the transition from the cosmic `dark age' \cite{Rees96} to a ionized 
universe populated with luminous sources. While one of the important  
questions recently emerged is the nature (starbursts or AGNs?) and redshift 
distribution of the ultraluminous sub-mm sources discovered by {\it SCUBA} 
\cite{Hu98}, \cite{B98}, \cite{Li98}, of perhaps equal interest is the 
possible existence of a large population of faint galaxies still undetected 
at high-$z$, as the color-selected ground-based and {\it Hubble Deep Field} 
(HDF) samples include only the brightest and bluest star-forming objects. 
In hierarchical clustering cosmogonies, 
high-$z$ dwarfs and/or mini-quasars (i.e. an early generation of stars and 
accreting black holes in dark matter halos with circular velocities $v_c\sim 
50\,\kms$) may actually be one of the main source of UV photons and 
heavy elements at early epochs \cite{MR98}, \cite{HL97}, \cite{HL98}. 
\begin{figure*}
\centerline{\epsfig{file=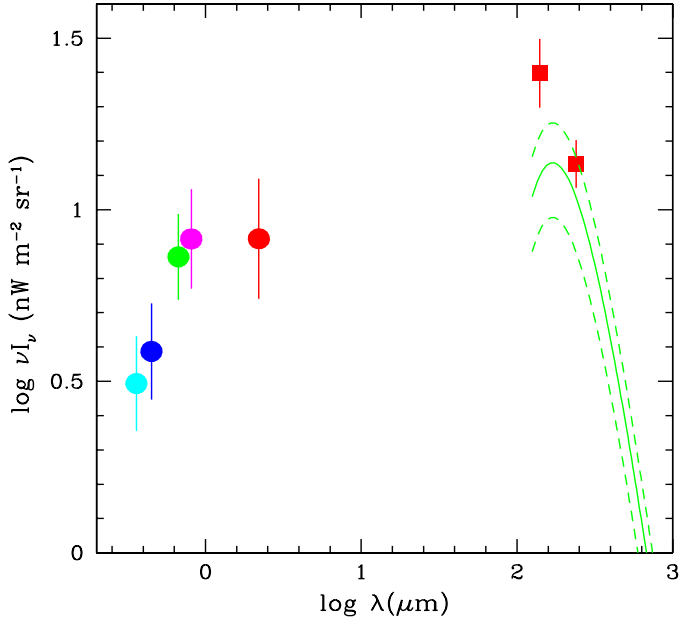,height=4.1in,width=4.3in}}
\vspace{-0.5cm}
\caption[h]{\small Spectrum of the extragalactic background light  
as derived from a compilation of ground-based and space-based galaxy counts in
the $U, B, V, I,$ and $K$-bands ({\it filled dots}), together with the FIRAS 
125--5000 $\mu$m ({\it solid and dashed lines}) and DIRBE 140 and 240 $\mu$m 
({\it filled squares}) detections.
\label{fig1}}
\end{figure*}
In this talk I will focus on some of the open issues and controversies 
surrounding our present understanding of the history of the conversion of 
cold gas into stars within galaxies, and of the evolution with cosmic time 
of the space density of luminous sources. An Einstein-de Sitter universe 
($q_0=0.5$) with 
$H_0=50h_{50}\,\kmsmpc$ will be adopted in the following. 

\section{Extragalactic background light}

The extragalactic background light (EBL) is an indicator of the total 
luminosity of the universe. It provides unique information on the evolution 
of cosmic structures at all epochs, as the cumulative emission from galactic 
systems and AGNs is expected to be recorded in this background. Figure 1 
shows the optical
EBL from known galaxies together with the recent {\it COBE} results.
The value derived by integrating the galaxy counts \cite{Pozz98} down to very
faint magnitude levels [because of the flattening at faint 
magnitudes of the $N(m)$ differential counts most of the contribution to the 
optical EBL comes from relatively bright galaxies] 
implies a lower limit to the EBL intensity in the 
0.3--2.2 $\mu$m interval of $I_{\rm opt}\approx 12\,\eblunits$. When combined 
with the FIRAS and DIRBE measurements 
($I_{\rm FIR}\approx 16\,\eblunits$ in the 125--5000 $\mu$m range), this gives
an observed EBL intensity in excess of $28\,\eblunits$. 
The correction factor needed to account for the residual emission in the 2.2
to 125 $\mu$m region is probably $\lta2$ \cite{Dwe98}. We shall see below 
how a population of dusty AGNs could make a significant 
contribution to the FIR background. In this talk I shall adopt a 
conservative reference value for the total EBL intensity associated 
with star formation activity over the entire history of the universe of 
$I_{\rm EBL}=40\,I_{40}\, \eblunits$.

\section{Star formation history}

\begin{figure*}
\centerline{\epsfig{file=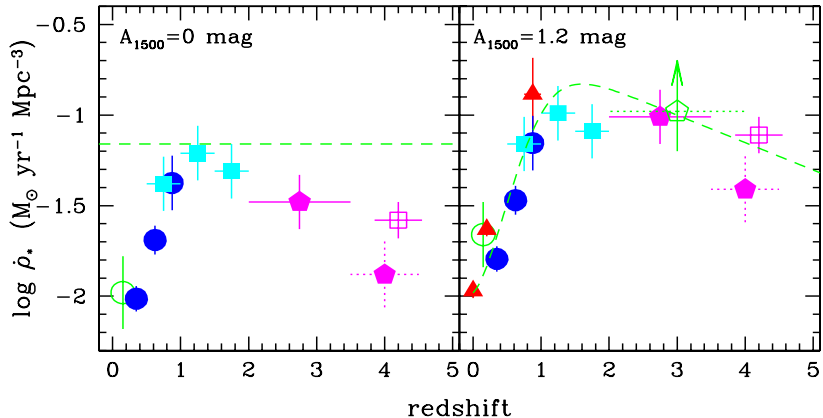,height=4.3in,width=4.5in}}
\vspace{-4.5cm}
\caption[h]{\small {\it Left}: Mean comoving density of star formation as 
a function of cosmic time. The data points with error bars have been inferred
from the UV-continuum luminosity densities of \cite{Li96} ({\it filled dots}),
\cite{Co97} ({\it filled squares}), \cite{M98} ({\it filled pentagons}), 
\cite{Treyer98} ({\it empty dot}), and \cite{Ste98} ({\it empty square}). 
The {\it dotted line} shows the fiducial rate, $\langle \dot{\rho_*}\rangle
=0.054\,\sfrd$, required to generate the observed EBL. {\it 
Right}: dust corrected values ($A_{1500}=1.2$ mag, SMC-type 
dust in a foreground screen). The H$\alpha$ determinations of \cite{Ga95}, 
\cite{TM98}, and \cite{Gla98} ({\it filled triangles}), together with the SCUBA 
lower limit \cite{Hu98} ({\it empty pentagon}) have been added for comparison.
\label{fig2}}
\end{figure*}
It has become familiar to interpret recent observations of high-redshift
sources via the comoving volume-averaged history of star formation.
This is the mean over cosmic time of the stochastic, possibly
short-lived star formation episodes of individual galaxies, and follows a
relatively simple dependence on redshift. Its latest version, uncorrected for 
dust extinction, is plotted in Figure 2 ({\it left panel}). The measurements 
are based upon
the rest-frame UV luminosity function (at 1500 and 2800 \AA), assumed
to be from young stellar populations \cite{M96}. The prescription for a 
`correct'
de-reddening of these values has been the subject of an ongoing debate.   
Dust may play a role in obscuring the UV continuum of Canada-France 
Reshift Survey (CFRS, $0.3<z<1$) and Lyman-break ($z\approx 3$) galaxies, 
as their colors are too red 
to be fit with an evolving stellar population and a Salpeter initial mass
function (IMF) \cite{M98}. The fiducial model of \cite{M98} had an upward 
correction factor of 1.4 at 2800 \AA, and 2.1 at 1500 \AA. Much larger corrections have been argued for by \cite{RR97}
($\times 10$ at $z=1$), \cite{Meu97} ($\times 15$ at $z=3$),
and \cite{Sa98} ($\times 16$ at $z>2$). As noted already by \cite{M96} and 
\cite{M98}, a consequence of such large extinction values is the possible 
overproduction of metals and red light at low redshifts. 
Most recently, the evidence for more moderate extinction
corrections has included measurements of star-formation rates (SFR) from Balmer lines 
by \cite{TM98} ($\times 2$ at $z=0.2$), \cite{Gla98} ($\times 3.1\pm 0.4$ at $z=1$),
and \cite{Max98} ($\times 2-6$ at $z=3$). {\it ISO} follow-up of CFRS fields \cite{F98} has shown that 
the star-formation density derived by FIR fluxes ($\times 2.3\pm 0.7$ at $0\le z\le 1$) is
about 3.5 times lower than in \cite{RR97}. Figure 2 ({\it right panel}) depicts an extinction-corrected  (with $A_{1500}=1.2$ mag, 0.4 mag higher than in 
\cite{M98}) version of the same plot. The best-fit cosmic star formation 
history (shown by the {\it dashed-line}) produces a total EBL of $37\,
\eblunits$. About 60\% of this is radiated in the UV$+$optical$+$near-IR
between 0.1 and 5 $\mu$m; the total amount of starlight that is 
absorbed by dust and reprocessed in the far-IR is $13\,\eblunits$. 
Because of the uncertainties associated with the incompleteness of  
the data sets, photometric redshift technique, dust reddening, and UV-to-SFR 
conversion, these numbers are only meant to be indicative. On the other
hand, the model is not in obvious disagreement with any of the observations, 
and is able, in particular, to provide a reasonable estimate of the 
near-IR luminosity density in the range $0\lta z\lta 1$ (Fig. 3).

\section{The stellar mass density today}

With the help of some simple stellar population synthesis tools it is 
possible at this stage to make an estimate of the integrated stellar mass
density today. The total {\it bolometric} luminosity of a simple stellar 
population (a single generation of coeval stars) having mass $M$ 
can be well approximated by a power-law with time for all ages $t\gta 100$ 
Myr, 
\begin{equation}
L(t)=1.3\,L_\odot {M\over M_\odot} \left({t\over 1\,{\rm Gyr}}\right)^{-0.8}
\end{equation}
(cf. \cite{Bu95}), 
where we have assumed solar metallicity and a Salpeter IMF truncated at 0.1
and 125 $M_\odot$. In a stellar system with arbitrary star-formation rate per
unit cosmological volume, $\dot \rho_*$, the comoving bolometric emissivity 
at time $t$ is given by the convolution integral
\begin{equation}
\rho_{\rm bol}(t)=\int_0^t L(\tau)\dot \rho_*(t-\tau)d\tau.
\end{equation}
The total background light observed at Earth ($t=t_H$) is 
\begin{equation}
I_{\rm EBL}={c\over 4\pi} \int_0^{t_H} {\rho_{\rm bol}(t)\over 1+z}dt,
\end{equation}
where the factor $(1+z)$ at the denominator is lost to cosmic expansion 
when converting from observed to radiated luminosity density. From the 
above equations it is easy to derive 
\begin{equation}
I_{\rm EBL}=740\,\eblunits \langle {\dot\rho_*\over \sfrd}\rangle 
\left({t_H\over 13\,{\rm Gyr}}\right)^{1.87}.
\end{equation}
The observations shown in Figure 1 therefore imply a ``fiducial'' mean star 
formation density of $\langle \dot\rho_*\rangle=0.054\, I_{40}\,\sfrd$. 
In the instantaneous recycling approximation, the total stellar mass 
density observed today is 
\begin{equation}
\rho_*(t_H)=(1-R)\int_0^{t_H} \dot \rho_*(t)dt\approx 5\times 10^8\,I_{40}\,
\mden, 
\end{equation}
(corresponding to $\Omega_*=0.007\,I_{40}$) where $R$ is the mass fraction 
of a generation of stars that is returned to the interstellar medium, 
$R\approx 0.3$ for a Salpeter IMF. The optical/{\it COBE} background 
therefore requires that about 10\% of the nucleosynthetic baryons ($\Omega_b
h_{50}=0.08$ \cite{Bur98}) are in the forms of stars and their remnants.
The predicted stellar mass-to-blue light ratio is $\langle 
M/L_B\rangle\approx 5$.
Note that these values are quite sensitive to the
lower-mass cutoff of the IMF, as very-low mass stars can contribute
significantly to the mass but not to the integrated light of the whole stellar
population. A lower cutoff of 0.5$\msun$ instead of the 0.1$\msun$ adopted
would decrease the mass-to-light ratio (and $\Omega_*$) by a factor of 1.9 
for a Salpeter function. 

%
\begin{figure*}
\noindent\begin{minipage}[b]{7.8truecm}
\epsfig{file=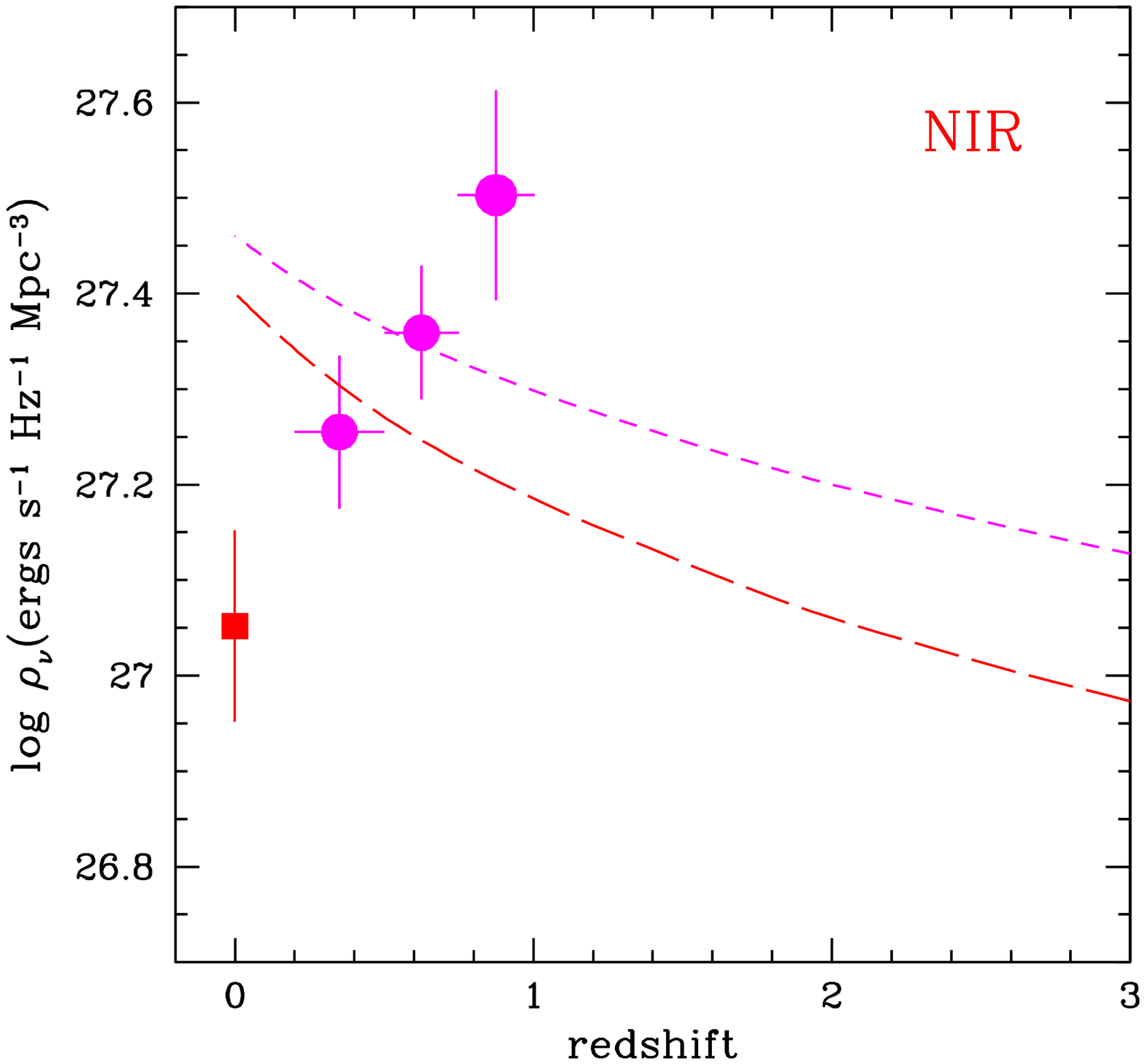,height=7.85truecm,width=7.85truecm}
\end{minipage}
\hfill
\noindent\begin{minipage}[b]{7.1truecm}
\epsfig{file=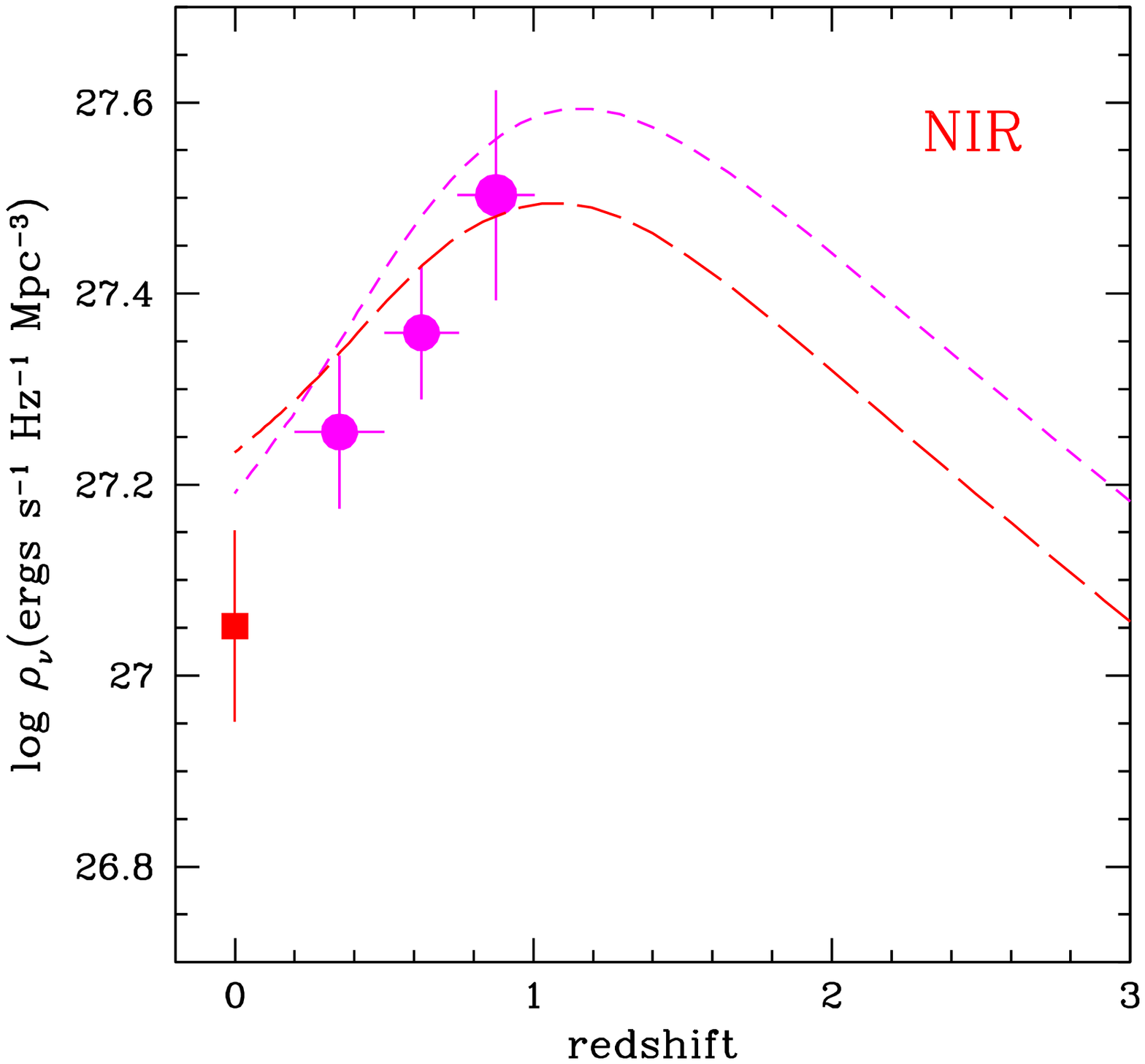,height=7.85truecm,width=7.85truecm}
\end{minipage}\hfill
\caption[h]{\small {\it Left}: Evolution of the near-IR luminosity density 
at rest-frame wavelengths of 1.0 $\mu$m ({\it long-dashed line}) and 2.2 
$\mu$m ({\it short-dashed line}). The data points are taken from \cite{Li96} 
({\it filled dots}) and \cite{Ga97} ({\it filled square}). 
The model assumes a constant star-formation rate of $\dot{\rho_*}=0.054\,
\sfrd$ (Salpeter IMF).  {\it Right}: Same but with the star-formation
history depicted in the right panel of Fig. 2. 
\label{fig3}}
\end{figure*}
\section{A constant star-formation density?}

Based on the agreement between the $z\approx 3$ and $z\approx 4$ luminosity 
functions at the bright end, it has been recently argued by \cite{Ste98} that 
the decline in the luminosity density of faint HDF Lyman-break galaxies 
observed in the same redshift interval \cite{M96} may not be real, but simply 
due to sample variance in the HDF. When extinction corrections are applied, 
the emissivity per unit comoving volume due to star formation may then 
remain essentially flat for all redshift $z\gta 1$ (see Fig. 2). While this
has obvious implications for hierarchical models of structure formation,
the epoch of first light, and the reionization of the intergalactic medium 
(IGM), it is also interesting to speculate on the possibility of a constant
star-formation density at {\it all} epochs $0\le z\le 5$, as recently 
advocated by \cite{Pasc98}. Figure 3 ({\it left panel}) shows the time 
evolution of the near-IR
rest-frame luminosity density of a stellar population characterized by a 
Salpeter IMF, solar metallicity, and a (constant) star-formation rate of 
$\dot{\rho_*}=0.054\,\sfrd$ (needed to produce the observed EBL).
\footnote{The near-IR light is dominated by near-solar mass evolved stars, 
the progenitors of which make up the bulk of a galaxy's stellar mass, and is 
sensitive to the past star-formation history.} The predicted evolution 
appears to be a poor match to the observations: it overpredicts the 
local $K$-band luminosity density \cite{Ga97} and undepredicts the 1$\,\mu$m 
emissivity at $z\approx 1$ from the CFRS survey \cite{Li96}. 

\section{ A population of hidden AGNs?}

Recent dynamical evidence indicates that supermassive black holes reside 
at the center of most nearby galaxies. The available data (about 30 objects) 
show a strong correlation (but with a large scatter) between bulge and black 
hole mass \cite{Mag98}, with $M_{\rm bh}=0.006 \, M_{\rm bulge}$ as a 
best-fit. The total mass density in spheroids today is $\Omega_{\rm bulge}=
0.0036^{+0.0024}_{-0.0017}$ \cite{Fuk98}, implying a mean mass density of
dead quasars 
\begin{equation}
\rho_{\rm bh}=1.34^{+0.9}_{-0.6}\times 10^6\,\mden.
\end{equation}
Noting that the observed energy density from all quasars is equal to the
emitted energy divided by the average quasar redshift \cite{Zol82}, the 
total contribution to the EBL from accretion onto black holes is 
\begin{equation}
I_{\rm bh}={c^3\over 4\pi} {\eta \rho_{\rm bh}\over \langle 1+z\rangle}
\approx 18\,\eblunits \eta_{0.1}\langle 1+z\rangle^{-1},
\end{equation}
where $\eta_{0.1}$ is the efficiency for transforming accreted rest-mass 
energy into radiation (in units of 10\%). A population of AGNs at (say) 
$z\sim 1$ could then make a significant contribution to the FIR background
if dust-obscured accretion onto supermassive black holes is an efficient
process \cite{Hae98}.
 
\section{Reionization of the IGM}

The history of the transition from a neutral universe to one that is almost
fully ionized can reveal the character of cosmological ionizing sources
and constrain the star formation activity at high redshifts.   
The existence of a filamentary, low-density intergalactic medium (IGM), which
contains the bulk of the hydrogen and helium in the universe, is predicted 
as a product of primordial nucleosynthesis \cite{CST} and of hierarchical 
models of gravitational instability with ``cold dark matter'' (CDM) 
\cite{C94}, \cite{H96}. The application of the Gunn-Peterson constraint on 
the amount of smoothly distributed neutral material along the line of sight 
to distant objects requires the hydrogen component of the diffuse IGM to 
have been highly ionized by $z\approx 5$ \cite{SSG}, and the helium 
component by $z\approx 2.5$ \cite{DKZ}. From QSO absorption studies we also 
know that neutral hydrogen accounts for only a small fraction, $\sim 10\%$, 
of the nucleosynthetic baryons at early epochs \cite{LWT}. It thus appears 
that substantial sources of ultraviolet photons were present at $z\gta 5$, 
perhaps low-luminosity quasars \cite{HL98} or a first generation of stars in 
virialized dark matter halos with $T_{\rm vir}\sim 10^4-10^{5} \,$K 
\cite{OG96}, \cite{HL97}, \cite{MR98}. 
\begin{figure*}
\noindent\begin{minipage}[b]{7.8truecm}
\epsfig{file=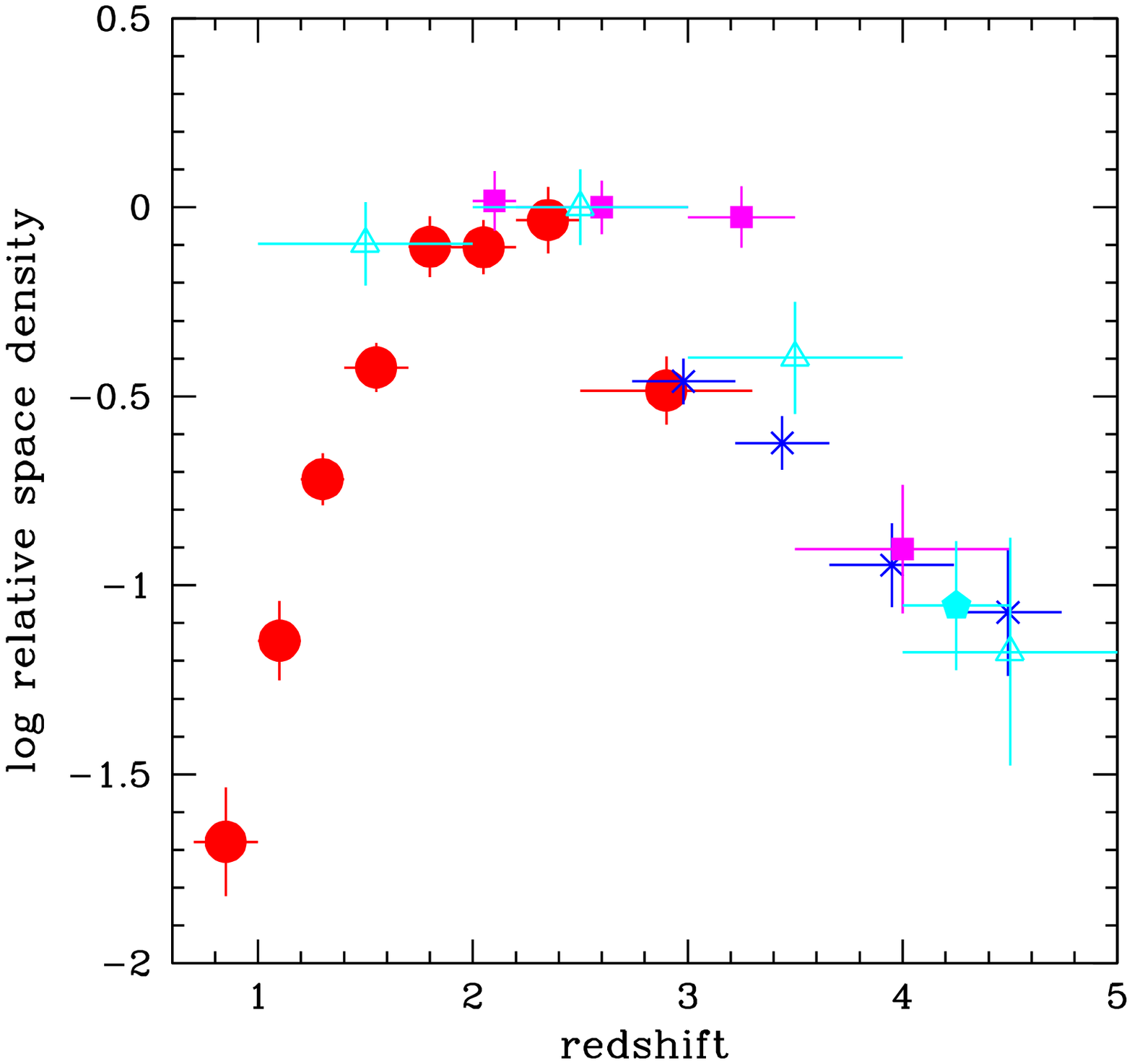,height=7.85truecm,width=7.85truecm}
\end{minipage}
\hfill
\noindent\begin{minipage}[b]{7.1truecm}
\epsfig{file=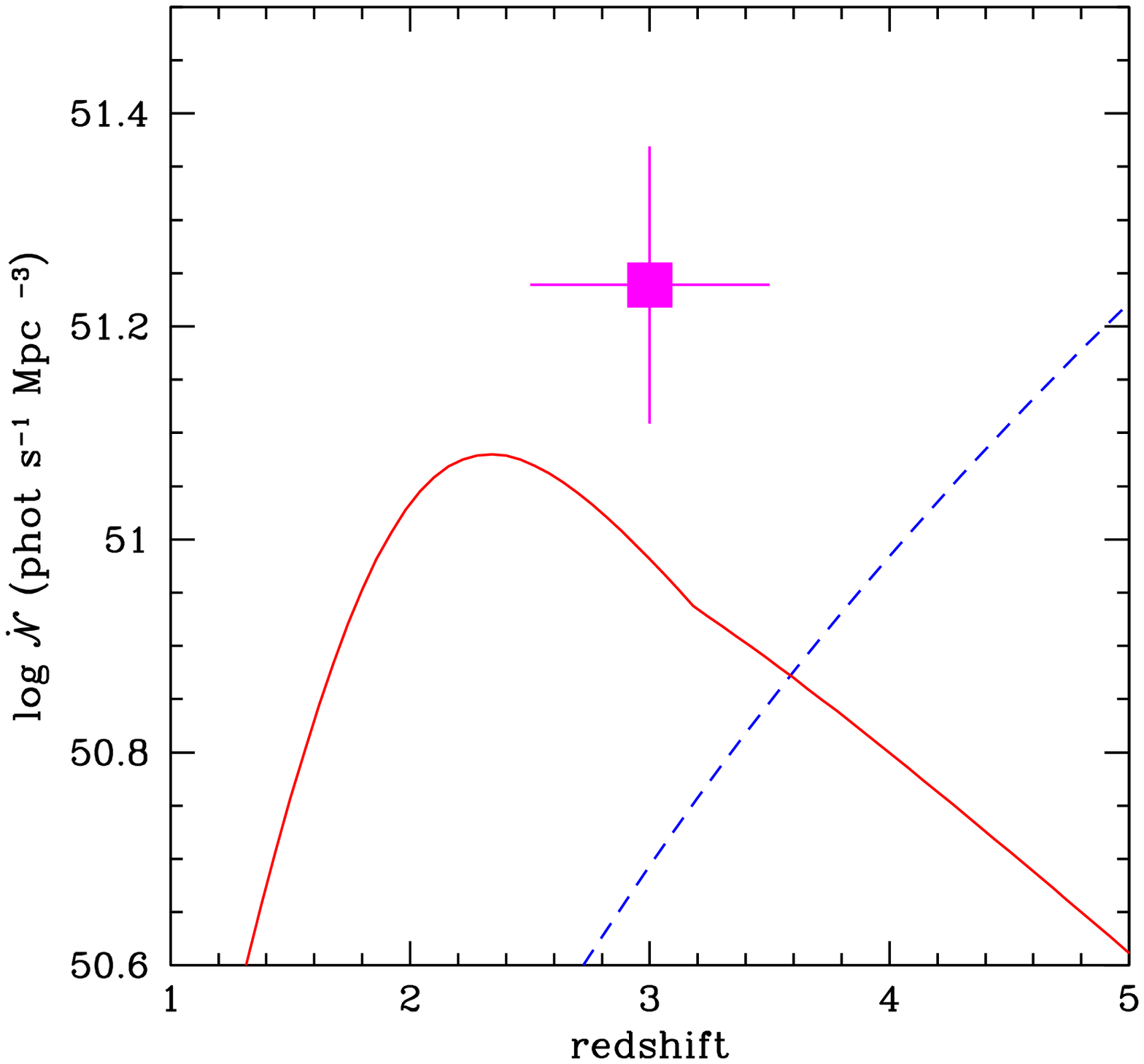,height=7.85truecm,width=7.85truecm}
\end{minipage}\hfill
\caption[h]{\small {\it Left}: comoving space density of bright QSOs as a 
function of redshift. The data points with error bars are taken from 
\cite{HS90} {\it (filled dots)}, \cite{WHO} {\it (filled squares)}, 
\cite{Sc95} {\it (crosses)}, and \cite{KDC} {\it (filled pentagon)}. 
The {\it empty triangles} show the space density of the Parkes flat-spectrum 
radio-loud quasars with $P>7.2\times 10^{26}\,$ W Hz$^{-1}$ sr$^{-1}$ 
\cite{H98}. {\it Right}: comoving emission rate of hydrogen Lyman-continuum 
photons ({\it solid line}) from QSOs, compared with the minimum rate 
({\it dashed line}) which is needed to fully ionize a fast recombining (with 
gas clumping factor $C=30$) Einstein--de Sitter universe with 
$\Omega_bh_{50}^2=0.08$. Models based on photoionization by quasar sources 
appear to fall short at $z=5$. 
The data point shows the estimated contribution of star-forming 
galaxies at $z\approx 3$, assuming that the fraction of Lyman continuum 
photons which escapes the galaxy \HI layers into the intergalactic medium 
is $f_{\rm esc}=0.5$ (see \cite{MHR} for details).
\label{fig4}}
\end{figure*}
The existence of a decline in the space density of bright quasars at redshifts
beyond $\sim 3$ was first suggested by \cite{O82}, and has been since then
the subject of a long-standing debate. In recent years, several optical 
surveys have consistently provided new evidence for a turnover in the QSO 
counts \cite{HS90}, \cite{WHO}, \cite{Sc95}, \cite{KDC}. 
The interpretation of the drop-off observed in optically selected samples is
equivocal, however, because of the possible bias introduced by dust 
obscuration arising from intervening systems. Radio emission, on the other 
hand, is unaffected by dust, and it has recently been shown \cite{Sha} that 
the space density of radio-loud quasars also decreases strongly for $z>3$. 
This argues that the turnover is indeed real and that dust along the line of 
sight has a minimal effect on optically-selected QSOs (Figure 4, {\it left
panel}). 
The QSO emission rate of hydrogen ionizing photons per unit comoving volume is
shown in Figure 4 ({\it right panel}) \cite{MHR}. It is important to 
notice that the procedure
adopted to derive this quantity implies a large correction for incompleteness
at high-$z$. With a fit to the quasar luminosity function (LF) which goes as 
$\phi(L)\propto L^{-1.64}$ at the faint end \cite{Pei95}, the
contribution to the emissivity converges rather slowly, as $L^{0.36}$. At
$z=4$, for example, the blue magnitude at the break of the LF is $M_*\approx 
-25.4$, comparable or slightly fainter than the limits of current high-$z$  
QSO surveys. A large fraction, about 90\% at $z=4$ and even higher at 
earlier epochs, of the ionizing quasar emissivity is therefore 
produced by sources that have not been actually observed, and are
assumed to be present based on an extrapolation from lower redshifts. 

Galaxies with ongoing star-formation are another obvious source of Lyman
continuum photons. Since the 
rest-frame UV continuum at 1500 \AA\ (redshifted into the visible band for a
source at $z\approx 3$) is dominated by the same short-lived, massive stars
which are responsible for the emission of photons shortward of the Lyman edge,
the needed conversion factor, about one ionizing photon every 10 photons at
1500 \AA, is fairly insensitive to the assumed IMF and is independent of the
galaxy history for $t\gg 10^7\,$ yr. Figure 4 shows the estimated 
Lyman-continuum luminosity density of galaxies at $z\approx 3$.\footnote{At 
all ages $\gta 0.1$ Gyr one has $L(1500)/L(912)\approx 6$ for a Salpeter mass 
function and constant SFR \cite{BC98}. This number neglects any correction 
for intrinsic \HI absorption.}~ The data point assumes a value of 
$f_{\rm esc}=0.5$ for the unknown fraction of ionizing photons which escapes 
the galaxy \HI layers into the intergalactic medium. 
A substantial population of dwarf galaxies below the detection threshold, 
i.e. having star-formation rates $<0.3\sfr$, and with a space density in 
excess of
that predicted by extrapolating to faint magnitudes the $\alpha=1.38$ best-fit
Schechter function, may be expected to form at early times in hierarchical 
clustering models, and has been recently proposed by \cite{MR98} and 
\cite{MHR} as a possible candidate for photoionizing the IGM at these 
epochs. One 
should note that, while highly reddened galaxies at high 
redshifts would be missed by the dropout color technique (which isolates 
sources that have blue colors in the optical and a sharp drop in the 
rest-frame UV), it seems unlikely that very dusty objects (with $f_{\rm esc}
\ll 1$) would contribute in any significant manner to the ionizing 
metagalactic flux. 

As the hydrogen mean recombination timescale, $\bar{t}_{\rm rec}$, at high 
redshifts is much smaller than the then Hubble time \cite{MHR}, it is 
possible to compute 
at any given epoch a critical value for the photon emission rate per unit 
cosmological comoving volume, 
\begin{equation}
\dot {\cal N}_{\rm ion}(z)={\bar{n}_\nH(0)\over \bar{t}_{\rm rec}(z)}=(10^{51.2}\,
\ndotunits)\, C_{30} \left({1+z\over 6}\right)^{3}\left({\Omega_b 
h_{50}^2\over 0.08}\right)^2, 
\label{eq:caln}
\end{equation}
independently of the (unknown) previous emission history of the universe: only
rates above  this value will provide enough UV photons to ionize the IGM by 
that epoch. Here $\bar{n}_\nH(0)$ is the mean hudrogen density of the 
expanding IGM at the present-epoch, and $C$ is the ionized hydrogen 
clumping factor. One can then compare our determinations of 
$\dot {\cal N}_{\rm
ion}$ to the estimated contribution from QSOs and star-forming galaxies. 
The uncertainty on this critical rate is difficult to estimate, as it depends 
on the clumpiness of the IGM  (scaled in the expression above 
to the value inferred at $z=5$ from numerical simulations \cite{GO97})
and the nucleosynthesis constrained baryon density. The 
evolution of the critical rate as a function of redshift is plotted in Figure 
4. While $\dot {\cal N}_{\rm ion}$ is comparable to the quasar 
contribution at 
$z\gta 3$, there is some indication of a deficit of Lyman 
continuum photons at $z=5$. For bright, massive galaxies to produce enough UV 
radiation at
$z=5$, their space density would have to be comparable to the one observed at
$z\approx 3$, with most ionizing photons being able to escape freely from the
regions of star formation into the IGM. This scenario may be in 
conflict with  direct observations of local starbursts below
the Lyman limit showing that at most a few percent of the stellar ionizing
radiation produced by these luminous sources actually escapes into the IGM 
\cite{Le95}.\footnote{Note that, at $z=3$, Lyman-break galaxies would radiate 
more ionizing photons than QSOs for $f_{\rm esc}\gta 30\%$.}~If, on the other 
hand, faint QSOs with (say) $M_\AB=-19$ at rest-frame ultraviolet frequencies 
were to provide {\it all} the required ionizing flux, their 
comoving space density would be such ($0.0015\,$Mpc$^{-3}$) that about 50 
of them would expected in the HDF down to $I_\AB=27.2$.  At $z\gta 5$, they 
would appear very red in $V-I$ as the \Lya forest is shifted into the visible.
This simple model can be ruled out, however, as there is only a handful (7) 
of sources in the HDF with $(V-I)_\AB>1.5$ mag down to this magnitude limit.
  
It is interesting to convert the derived value of $\dot {\cal N}_{\rm ion}$  
into a ``minimum'' SFR per unit (comoving) volume, $\dot 
\rho_*$ (hereafter we assume $\Omega_bh_{50}^2=0.08$ and $C=30$):
\begin{equation}
{\dot \rho_*}(z)=\dot {\cal N}_{\rm ion}(z) \times 10^{-53.1} f_{\rm esc}^{-1}
\approx 0.013 f_{\rm esc}^{-1} \left({1+z\over 6}\right)^3\ \sfrd. \label{eq:sfr} 
\end{equation}  
(The conversion factor assumes a Salpeter IMF with solar metallicity). The 
star-formation density given in
equation (\ref{eq:sfr}) is comparable with the value directly ``observed'' 
(i.e., uncorrected for dust reddening) at $z\approx 3$ \cite{M98}.

\vfill
\end{document}